\documentclass[12pt]{article}
\usepackage{graphicx}
\newcommand{\be}{\begin{equation}}
\newcommand{\bea}{\begin{eqnarray} \nonumber}
\newcommand{\ee}{\end{equation}}
\newcommand{\eea}{\end{eqnarray}}

 \def\(({\left(}
 \def\)){\right)}
\def\[[{\left[}
\def\]]{\right]}

\def \form#1 {eq. (\ref{#1}) }
\def \parziale#1#2  {{\partial {#1} \over \partial {#2}}}

\def \cP{{\cal P}}

\def \ba#1 {\overline{#1}}

\begin{document}

\title{On the survey-propagation equations for the random K-satisfiability problem}
\author{ Giorgio Parisi\\
Dipartimento di Fisica, Sezione INFN, SMC and
 UdRm1 of INFM,\\
Universit\`a di Roma ``La Sapienza'', \\
Piazzale Aldo Moro 2,
I-00185 Rome (Italy)}

\maketitle

\
\begin{abstract}

\noindent In this note we study  the existence of a solution to the survey-propagation equations for the
random K-satisfiability problem for a given instance. We conjecture that when the number of variables goes to infinity, 
the solution of these equations for a given instance can be approximated by the solution of the corresponding equations 
on an infinite tree. We conjecture (and we bring numerical evidence) that the survey-propagation equations on the 
infinite tree have an unique solution in the suitable range of parameters.
 
\end{abstract}

\section{Introduction}

Recently many progresses \cite{MPZ,MZ} have been done on the analytic and numerical study of the random K-satisfiability
problem \cite{COOK,KS,sat,sat0}, using the approach of survey-propagation that generalizes the more old approach
based on the ``Min-Sum'' \footnote{The ``Min-Sum'' is the the zero temperature limit of the ``Sum-Product'' algorithm
and sometimes is also called belief propagation.  In the statistical mechanics language \cite{MPV} the belief
propagation equations are the extension of the TAP equations for spin glasses \cite{TAP} and the survey-propagation equations are
the TAP equations generalized to the broken replica case.} algorithm \cite{BP,MPV,factor,MoZ} .

The aim of this note is to sketch a possible path to a proof of the existence an uniqueness of the survey-propagation
equations for a given instance of the problem.  Before presenting the main arguments, for reader convenience I will
present an heuristic derivation of the survey-propagation equations in section 2; the full derivation can be found in the
original papers \cite{MPZ,MZ,primo,MP1,MP2}.  In section 3, I will present the main sequence of conjectures that may lead to
the proof of the existence and uniqueness of the survey-propagation equations in the appropriate range of parameters. 
Finally I will present some conclusions.

\section{A fast heuristic derivation of the survey equations}
\subsection{The random K-sat problem }
In the random K-sat problem there are  $N$ variable $\sigma(i)$ that may be true of false (the index $i$ will sometime 
called  a node).  An instance of the 
problem is given by a set of $M\equiv \alpha N$ clauses.  Each clause is characterized by set of three nodes 
($i_{1}$,$i_{2}$, $i_{3}$), that belong to the interval $1-N$ and by three Boolean variables ($b_{1}$,$b_{2}$, $b_{3}$).
In the random case the $i$ and $b$ variables are random with flat probability distribution.
Each clause $c$ is true if the expression 
\be
E_{c}\equiv (\sigma(i^{c}_{1})\ XOR\ b^{c}_{1}) \ OR \ (\sigma(i^{c}_{2})\ XOR\ b^{c}_{2}) \ OR \ 
(\sigma(i^{c}_{3})\ XOR\ b^{c}_{3})
\ee
is true.
The problem is satisfiable iff we can find a set  of the variables $\sigma$ such that all the clauses are true. The 
entropy \cite{MoZ} of a 
satisfiable problem is the logarithm of the number of the different sets of   the $\sigma$  variables that make all the clauses  true.

To a given problem we can associate a graph (the factor graph \cite{factor}) where the nodes are connected to the 
clauses ($3\alpha$ in average) and each clause is connected to three nodes.  The properties of this graph play a 
very important role.

The goal of the analytic approach consists in finding for given $\alpha$ and for large values of $N$ the probability
that a random problem (i.e. a problem with random chosen clauses) is satisfiable. The $0-1$ law
\cite{KS,sat0,01} is supposed to be valid: for $\alpha<\alpha_{c}$ all systems (with probability one when $N$ goes to infinity) are
satisfiable and their entropy is proportional to $N$ with a constant of proportionality that does not depend on the
problem.  On the other hand, for $\alpha>\alpha_{c}$ no random system (with probability one) is satisfiable.  An
heuristic argument has been given \cite{MPZ,MZ} that suggest that $\alpha_{c}=\alpha ^{*}\approx 4.27$ where
$\alpha^{*}$ can be computed using the survey-propagation equations defined later.  There is already an incomplete proof
\cite{FL} that $\alpha ^{*}$ is a rigorous upper bound to $\alpha _{c}$.

\subsection{The belief propagation equations}
In the following analysis it will 
be crucial that in limit $N \to \infty$ the problem becomes locally 
trivial: this   makes possible the computations of $\alpha_{c}$ and of the other properties of the system.
Let us be more precise. We can define a distance 
among the $N$ nodes in the following way:
\begin{itemize}
    \item Two nodes are at distance 1 if there is a clause that contains both of them. At large $N$ the number of nodes 
    at distance 1 from a given node  has a Poisson distribution with average $6 \alpha$.
    \item Two nodes are at distance $k$ if they are not at a distance $k-1$ and there is a chain of $k$ overlapping 
    clauses that touch both of them (or equivalently the second node is at distance 1 from a node at distance $k-1$ 
    from the first node).
\end{itemize}
In the limit $N \to \infty$ the set of nodes at distance $k$ from a given node form a tree,
without closed loops: locally the system looks like a tree.  The solution of the K-sat problem on a tree (with given
boundary conditions) can be trivially done: the belief propagation algorithm (defined later) is exact.  When $N$ goes to
infinity the systems is not a tree (with probability 1) and this fact makes hard to find an actual solution of the
problem for a given instance; it may even destroy the global existence of a solution to the belief propagation
equations.  However the local treeness of the problem is enough to allow an analytic treatment.

Let us take a large system for $\alpha<\alpha_{c}$ and let us consider the set $C$ of all configurations that  satisfies 
all the clauses. Our first task is  to compute the probabilities $p_T(i)$ and $p_F(i)$ that the variable 
$\sigma(i)$ is true or false (obviously $p_T(i)+p_F(i)=1$), if $\sigma(i)$ belong to a random configuration in $C$.

The presence of a simple local structure allows us to write down simple local equations \cite{primo,MP1,MP2}.
In the case of belief
propagation equation \cite{BP} one proceed as follows.  For each clause that contains the node $i$ (we will use the
notation $c \in i$ although it may be not the most appropriate)  $p_T(i,c)$ is the probability
that the variable $\sigma(i)$ would be true in absence of the clause $c$.  If the node $i^{c}_{1}$ were contained in only one
clause, we would have that 
\bea p_T(i^{c}_{1})=
u_T(p_T(i^{c}_{2},c),p_T(i^{c}_{3},c),b^{c}_{1},b^{c}_{2},b^{c}_{3}) \equiv u_T(i,c) \ , \\
p_F(i^{c}_{1})=1- u_T(i,c) \ ,
\eea
where $u_T$ is an appropriate function that is defined by the previous relation.
An easy computation shows that when all the $b$ are false, the variable $\sigma(i^{c}_{1})$ must be true if  both  
variable  $\sigma(i^{c}_{2})$ and  $\sigma(i^{c}_{3})$ are false, otherwise it can have any value. Therefore we have in 
this case that
\be
u_T(i,c)=p_F(i^{c}_{2},c) p_F(i^{c}_{3},c)+\frac {1-p_F(i^{c}_{2},c) p_F(i^{c}_{3},c)}{2}
\ee
In a similar way, if some of the $b$ variable are true, we should exchange the $p_F$ with the $p_T$ for the corresponding 
variables. Finally we have that 
\bea
p_T(i,c)={\prod_{d\in i, d\ne c}u_T(i,d) \over Z(i,c)} \\
p_F(i,c)={\prod_{d\in i, d\ne c}u_F(i,d) \over Z(i,c)}  \nonumber \\
Z(i,c)=\prod_{d\in i, d\ne c}u_T(i,d)+\prod_{d\in i, d\ne c}u_F(i,d).
\eea

In total there are $3M$ variables $p_T(i,c)$ and $3M$ equations.  These equations are called in the literature under 
different name (e.g. belief propagation, TAP equations \cite{TAP}) and we naively expect that these equations (belief
propagations) are satisfied (or quasi-satisfied, i. e. corrections of order $1/N$ can be present \cite{MP2}).  For the
time being let us suppose that such a solution  exists and it is unique.  In such case we expect that 

\bea
p_T(i)={\prod_{d\in i }u_T(i,d) \over Z(i)} \\
p_F(i)={\prod_{d\in i, }u_F(i) \over Z(i)}  \nonumber \\
Z(i)=\prod_{d\in i, }u_T(i,d)+\prod_{d\in i}u_F(i,d).
\eea
We note  the previous formulae can be written in a more compact way 
 if we introduce a two dimensional vector $\vec{p}$, with components $p_T$ and $p_F$.
We define the product of these vector 
\be
c_T=a_T \  b_T \ \ \ \ \ c_F=a_F \  b_F , 
\ee
if $\vec{c}=\vec{a} \cdot \vec{b}$. 

If the norm of a vector is defined by
\be
|\vec{a}|=a_T+a_F \ ,
\ee
we finally find that
\be
p(i,c)={\prod_{d\in i, d\ne c}u_T(i,d) \over |\prod_{d\in i, d\ne c}u_T(i,d)|}\ .
\ee

For each clause $c$ we can define the probability that the clause  would be satisfied  in a system where the clause is 
not present. We will denote this probability by $Z(c)$. One finds that in the case where all the $b$ variables are false 
\be
Z(c)=1-p_F(i^{c}_{1})p_F(i^{c}_{2})p_F(i^{c}_{2}) \ .
\ee
Finally the total entropy (apart correction that are subleading when $N$ goes to infinity) is given by 
\be
S=-\sum_{i} \log(Z(i)) +2\sum_{c}\log(Z(c)) \ .
\ee

\subsection{The survey propagation equations}

One can argue that the situation is more complex \cite{MPV,MP1,MP2}.  The belief propagation algorithm works at 
low value of $\alpha$ but it fails when $\alpha$ becomes too large. The previous  equations may have multiple
solutions or quasi-solutions that are very different one from the other.  

This fact has been interpreted \cite{MPZ,MZ,MP1,MP2} in the following way: the set of all configurations $C$ that
satisfy all the formulae can be divided into many sets that are well separated one by the others (this sets are 
sometimes called states in statistical mechanics \cite{MPV} or lumps \cite{TALE}).  The previous belief
equations correspond to the probability restricted to one given set.  

The new picture is the following: we have an
exponential large number of solutions (or quasi-solutions) of the belief equations (we call this set $B$) and we would
like to know this number (i.e. the exponential of the complexity $\Sigma(\alpha)$).  The 
total number of configurations that satisfies all the clauses is given by
\be
\exp(S(\alpha))=\exp(\Sigma(\alpha)+S_{s}(\alpha))
\ee
where $\exp(S_{s}(\alpha))$ is the total number of configurations that satisfy all the clauses in a generic state.

One expects that $\Sigma(\alpha)$ vanishes at $\alpha_{c}$, so that it computation is extremely important.  In
order to reach this goal we can mimic the steps that we have done for going  from the variables  $\sigma$ to the probability
$p_T(i)$; however in this case we are going to play the game at an higher level of abstractions, where states (or
quasi-solutions of the belief equations) play the same role of a single configuration of the $\sigma(i)$ in the previous
approach.

The new quantity is the full survey probability $s(i|  \vec{p})$, i.e. a function of $\vec{p}$ that is defined at each node
that is equal to the probability of finding a solution (or a quasi-solution) of the belief equations with $\vec{p}(i)=\vec{p}$.  In
other words $s(i| \vec{p})$ an indirect probability, i.e. a probability of a probability. 

We  introduce the quantity $s(i,c| \vec{p})$ that is the distribution probability
for the probability (i.e. $s(i|\vec{p})$) when the clause $c$ is removed.  The equations for the full survey probability
$s(i,c|\vec{p})$ can be obtained using the techniques of \cite{MP1,MP2}, however we will not consider them here.  Indeed we
need them if our aim is to compute the total entropy (or $S_{s}(\alpha)$).  If our aim is more
modest and we want to compute only $\Sigma(\alpha)$ and consequently $\alpha_{c}$, a simple approach is possible
\cite{MPZ,MZ,MP2}.

The crucial observation is that a given solution of the belief equations may
have $ p_T(i,c)=1$  or $ p_T(i,c)=0$ or $0<
p_T(i,c)<1$.  

We can coarse grain the probability distribution of the beliefs by introducing the quantity $s_T(i,c)$
that is defined as the probability of finding $ p_T(i,c)=1$, in the same way $s_F(i,c)$ as the probability of finding $
p_T(i,c)=0$ and $s_{I}(i,c)$ is the probability of finding $0< p_T(i,c)<1$.  As discussed in \cite{MP2,MPZ,MZ}, by
considering the equations for only these coarse grained surveys, it is possible to compute the complexity $\Sigma$ and
the value of $\alpha_{c}$.  

We can use a more compact notation by introducing a three dimensional vector $\vec{s}$ given
by

\be
\vec{s}= \{ s_T,s_I,s_F\} \ .
\ee
Everything works as before with the only difference that we have a three component vector instead of a two component 
vector. 
Generalizing the previous arguments one can introduce the quantities $\vec{u}(i,c)$ that is the value that the 
survey at $i$ would take if only the clause $c$ would be present in $i$.
In the case where all the $b$ are false, a simple computation gives 
\be
\vec{u}(i,c) =\{ s_F(i^{c}_{2},c), s_F(i^{c}_{3},c),\   1- s_F(i^{c}_{2},c), s_F(i^{c}_{3},c) ,\  0 \} \ . \label{A}
\ee
The formula can generalized as before \footnote{It always happens that the  vector $\vec{u}$ has only one  zero 
component ($u_{T}u_{F}=0$). This fact may be used to further simplify the analysis.} to the case  of different values of $b$. 
One finally finds
\be
\vec{p}(i,c)={\prod_{d\in i, d\ne c}\vec{u}(i,d) \over |\prod_{d\in i, d\ne c}\vec{u}(i,d)|} \ ,
\ee
where we have defined product in such a way that 
\be
\vec{a}\vec{b}=\{a_T b_T+a_{I}  b_T  +a_T b_{I}, a_{I}  b_{I},  \ a_F\ b_F+a_{I}\  b_F  +a_F\  b_{I} \}
		 . \label{B}
\ee
The previous equations are the survey propagation equations 
(as defined in \cite{MPZ,MZ}) and the reader can find there the details of the derivation.

If one finds a solution to the survey equation  one can compute the survey 
probability as 
\be
\vec{s}(i)={\prod_{d\in i,}\vec{u}(i,d) \over | \prod_{d\in i} \vec{u}(i,d)| }
\ee
and the total complexity is given by
\be
\Sigma=-\sum_{i} \log(Z(i) +2\sum_{c}\log(Z(c))
\ee
where now
\be
Z(i)=\ln(|\prod_{d\in i}\vec{u}(i,d)|), \ \ \ 
Z(c)=\ln(|\vec{s}(i,c)\vec{u}(i,c)|)
\ee

\section{The infinite rooted tree and some conjectures}
In the previous section we have sketched an heuristic derivation of the survey equations: it should be clear that these
equations should be taken as they are and  there is no warranty of any kind, either
expressed or implied, that the belief equations (that have been used heuristically to construct the survey equations) do have a
global solution.  
 \begin{figure} \begin{center}    
      \includegraphics[width=0.40\textwidth]{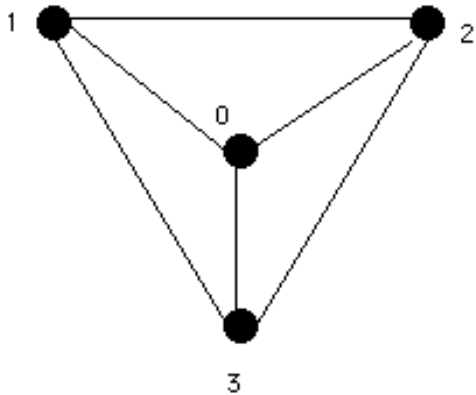}
      \end{center} \caption{ \label{AA}
 An example of a graph with loops.
}\end{figure}

The aim of this note is not to present a more precise derivation of the survey equation, but to discuss the problem of
finding solutions to the survey propagation equations: this is a well posed mathematical problem independently from the
origin of these equations.  If the survey equations would have no solutions, the previous arguments
would be empty or if survey equations had an exponential number of solutions, we should go up to an higher level of
abstraction.

Numerical experiments on systems of size up to $N= 10^{6}$ show that in the interval $\alpha_{L}< \alpha< \alpha_{U}$ 
($\alpha_{L}$ and $\alpha_{U}$ asymptotically do not depend on the system, they  are near to 3.9 and 4.36 for large $N$), 
the survey equations do have one solution that is obtained by iterations.  For $\alpha<\alpha_{L}$ the survey equation 
converge to the trivial solution $s_I(i,c)=1$.  On the other end for $\alpha >\alpha_{U}$ the iterative equations do not 
converge.  Moreover the difference among a solution for the survey propagation equations and those for a perturbed 
survey propagation equation (e.g. by adding or removing a clause, or by fixing a survey to an arbitrary value) diverges 
when we approach $\alpha_{U}$ from below.  The complexity change sign at a value $\alpha^{*}$ such that 
$\alpha_{U}>\alpha^{*}>\alpha_{L}$.
\subsection{The infinite tree}
These results call for an analytic proof.  The aim of this note is to suggest a possible approach.  

\begin{figure} \begin{center}    
      \includegraphics[width=0.55\textwidth]{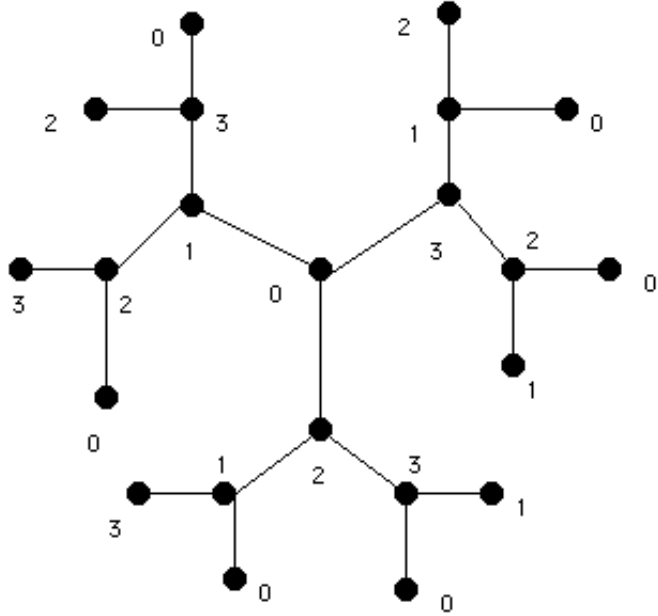}
      \end{center} \caption{ \label{C}
The central part of the infinite tree associated to the example shown in  fig. (\ref{AA}).
The numbers near the nodes are the values 
of the function $M$ evaluated at that node. }\end{figure}

We propose to generalize the construction that Aldous has recently used in the study of the random matching problem 
\cite{ALDOUS}.  For any node $i$ of a given problem for finite $N$ we associate an {\sl infinite} tree rooted in $i$ 
that is constructed in the following way \footnote{Sometimes the tree is not infinite, e.g. if the site $i$ has no 
neighbour, however for not too small $\alpha$ the tree is infinite in most of the cases.  If no loops were present (a 
rather unlikely possibility for large $N$) in the original graph the infinite rooted tree would be identical to the 
original graph.}.  
Let us denote by $\gamma$ a node of the infinite tree.  We assume that there is a function $M(\gamma)$ that maps the 
nodes of the infinite tree onto the nodes of the original problems in such a way that if $\gamma_{1}$, $\gamma_{2}$ and 
$\gamma_{3}$ belong to the same clause, also the three nodes $i_{k}=M(\gamma_{k})$ belong to the clause, (the 
variables $b_{k}$ of the two clauses have the same values).  We can further impose that the number of nodes at distance 
one from $\gamma$ is equal to the number of nodes at distance one from $M(\gamma)$.  The construction is simpler that it 
may looks.  In the cases of a problem with four variables and clauses that involve only two sites (i.e. a 2-sat, not 
3-sat for graphical convenience) the original graph is shown in figure \ref{AA}, while the center of the infinite graph, 
rooted in 0, is shown in fig.  (\ref{C}).

The construction of the infinite rooted tree is the simplest way to project a finite graph on a tree preserving as much 
as possible the structure of the original graph.  The fact that for large $N$ the original problem does not contain 
small loops implies that the original problem is locally very similar to a subset of the infinite tree.  This fact 
suggest that the solution on the K-sat problem on the infinite tree may give us information on the solution of the K-sat 
problem on the original problem.

An other object that we can construct is a random infinite tree.  It is a random tree where the number of nearest nodes 
has a Poisson distribution.  It is evident that for large $N$ the infinite tree associated to a given random problem 
becomes locally very near to a random infinite tree: the first contains correlations that vanish when $N$ goes to 
infinity.  Moreover the properties the infinite random tree can be studied analytically.  One would like to compare the 
properties of a given problem with the properties of the associated infinite tree; the hope is that the infinite tree 
associated to a random problem with $N$ clauses should become similar to the random infinite tree when $N$ goes to 
infinity.

We will say that a problem on the infinite tree has a unique solution iff, when we impose a generic boundary conditions one 
the $k+1$ shell, the behaviour at the center of the tree does not depend on the boundary conditions with probability one 
when $k$ goes to infinity.

The problem of computing a solution that satisfies all the clauses cannot have a unique solution in the previous sense:
 the entropy has a term proportional to $N$ and there is an exponentially large number of different solutions. 
 
The belief propagation equations have an unique solution for a given belief on the boundary (quasi solutions fade away 
if closed loops are absent).  An explicit computation show that when $\alpha$ is greater than a critical value (around 
3.9) \cite{MPZ,MZ} the solution of the belief does depend on the boundary.  One could also argue that if the solution of 
the belief equations would be independent from the boundary conditions when $k$ goes to infinity, there should be an 
essential unique solution of the belief equation for a given problem and for sufficiently large $\alpha$ this does not 
happen \cite{MPZ,MZ} (i.e. Aldous essential uniqueness property fails for the belief equations).

\subsection{Three conjectures}

We have now to exclude  the possibility that the survey propagation equations do have a stable solution on the rooted tree  
near $\alpha_{c}$. This is an highly non-trivial requirement that fails in other models or for other forms of survey 
equations in the K-sat problem.

If the approach of \cite{MPZ,MZ} is correct  the following conjectures should be true:
\begin{enumerate}
    
     \item For $\alpha <\alpha_{U}$ and large $N$ the infinite tree associated to random problem has one stable solution
     with probability one.  If this happens, the corresponding survey probabilities will be denoted by
     $\vec{s}_{\infty}(i,c)$.

    \item For large $N$ the survey probabilities $\vec{s}_{\infty}(i,c)$, should be an approximate solution of the 
    survey probability equations of the original problem.  In other words for a given sample and $\alpha <\alpha_{U}$ 
    there is solution of the survey probability equations near to $\vec{s}_{\infty}(i,c)$.

    \item The value of $\alpha_{U}$ is determined by the following condition: for $\alpha <\alpha_{U}$ the infinite 
    random tree has only one stable solution, while this does not happens $\alpha >\alpha_{U}$.
\end{enumerate}

A direct numerical test of these conjectures is not easy especially in the interesting region where $\alpha$ is near to 
4.  In principle it is possible to construct in an explicit way the first $k$ shells of the infinite tree and to study 
what happens for large $k$.  Unfortunately the number of first neighbour nodes is $6 \alpha$ so that for $\alpha=4$ the 
$k^{th}$ shell contains of the order $24^{k}$ nodes, that is a very large number for numerical analysis already for 
$k=5$.  If $N$ is not much greater than $24^{5}$, there will be many repetitions of the same nodes in the first five 
shells of the tree and on such scale the original problem does not looks very tree like.  I have done studied 
numerically problems up to $N=10^{6}$ and $k=4$ and the data (e.g. at $\alpha=4$) are consistent with the first two 
conjectures although it is difficult to arrive to convincing evidence.

If we accept also the last conjecture, we can compute the value of $\alpha_{U}$ on the infinite random 
lattice and we can compare it with the numerical  estimates for a given problem, i.e. $\alpha_{U}\approx 4.36$.
At this end we must  study the survey equations on  the infinite random tree.  In this case it is natural to
suppose a translational invariance property \cite{MPV,MP1,ALDOUS}.  

Let us call $\cP_{k}(\vec{s})$ the probability distribution of the survey in a generic node at distance $k$ form the
origine \footnote{For each node $i,$ we have to consider all the different surveys with one of the clauses removed, for
simplicity we will not indicate the $c$ dependence and in the following $\vec{s}(i)$ will be nickname for all the
$\vec{s}(i,c$).}. Translational invariance implies that $\cP_{k}(\vec{s})$ does not depend on $k$: it will be denoted by
$\cP(\vec{s})$.  This quantity plays a crucial role in the approach  \footnote {It is convenient to recall the heuristic
definition.of $\cP(\vec{s})$.  
We decompose into states the set of the configuration that satisfies all the conditions.  For each state we compute the 
belief probability that a given variable is true.  The 
survey probability  characterizes the distribution probability of the belief at a given site: the survey is a 
probability of a probability (an indirect probability).  Finally $\cP(\vec{s})$ is the probability of finding a site 
with that particular survey probability.  In other words $\cP(\vec{s})$ is a probability of a probability of a 
probability; in the simplifying case we are studying here is a function (we only care if a belief is equal to $\pm 1 $ 
or not), while in the more general case it would be a functional.}.

\subsection{A consistency check}

In principle it possible that this construction fails: the equations for the survey  may have a solution that depends 
on the boundary condition. In such a case a more complex construction should be done \cite{MP1} and the aim of this note 
is to exclude that this  happens.

The properties of $\cP(\vec{s})$ can be well studied numerically and hopefully analytically.  Indeed the surveys of the 
nodes on a shell at distance $k$ from the origine can be expressed in terms of the surveys at the nodes at 
distance $k+1$; using the supposed translational invariance of the probability distribution we get a consistency 
equation.  

The procedure is the following.  We consider a node with $z$ clauses, where $z$ has a Poisson distribution with average
$3 \alpha$ and we assume that the nearest nodes have the survey probability distributed according to $\cP(\vec{s})$ and we
compute the survey probabilities for the new node.  If we average on $z$ and on the random clauses we get a new survey
probability $\tilde{\cP}(\vec{s})$, that obviously depends on $\cP(\vec{s})$.  The equation for $\cP(\vec{s})$ are
simply 

\be \tilde{\cP}(\vec{s})=\cP(\vec{s}) \ .\label {FINAL} \ee 
Using the techniques of \cite{MPZ,MZ,MP1,MP2} one can also construct a functional $\Phi(\cP)$ such that the actual
solution of the equation (\ref{FINAL}) can be found by minimizing this functional (however we will not discuss this point).

It is a standard conjecture \cite{MP1} that equations of the form (\ref{FINAL}) can be solved by iteration (and this 
likely follows from the convexity properties of $\Phi(\cP)$).  If this is the case, we can use the method of population 
dynamics to find the solution of equation (\ref{FINAL}).  The method is very simple and can be trivially implemented on 
a computer.  It consists in describing a probability $P$ by an ensample of $L$ elements distributed according to this 
probability; the method becomes exact when $L$ goes to infinity.

We consider a set of $L$ surveys.  Starting from this set we generate a new ensemble of $L$ surveys by using the
standard procedure described in \cite{MPZ,MZ,MP1,MP2}.  The construction of an element of the new ensample is done as
follow. We extract a Poisson distributed integer $z$ with average $3 \alpha $.  We take $6 z$ surveys extracted randomly
from the $L$ surveys and we extract random the $b$ of the corresponding $k$ clauses.  Using equations
(\ref{A},\ref{B})  we compute one of the surveys that belong the new ensemble.  Finally by repeating this operation $L$
times we obtain the new ensample.  

By iterating this procedure $I$ time  we find for large $I$ a
probability distribution on the surveys, that is $I$ dependent.  This procedure can be done also for large values of $L$
(e.g. £$L=10^{6}$) and the convergence is rather fast (the corrections seem to be proportional to $1/L$ for generic
$\alpha$).  If the limits $L$ to $\infty$ and $I$ are smooth (the first should be done firstly) the resulting
probability distribution is a solution of the equation (\ref{FINAL}).

In the same approach we can ask what happens in the population dynamics if we start from two different sets of surveys at
the initial step.  Let us indicate the $i^{th}$ survey at the iteration $t$ with $\vec{s}(i,t)$.  Let assume to run
twice the population algorithm (with the same random number generator) but taking two different sets as starting points:
$\vec{s}_{1}(i,0)$ and $\vec{s}_{2}(i,0)$.  We expect that for large $i$ the probability distribution of the survey
should be the same; however it is a not evident if for a given $i$ 

\be \vec{s}_{1}(i,t)-\vec{s}_{2}(i,t)\to_{t \to \infty} 0  \label{BOH} \ .\ee 

It is natural to conjecture that if this happens, the survey equations have an unique solution
on a random infinite tree.  Indeed the computation of the surveys on the shell $M-I$ as function of the survey on the
shell $M-I+1$, can be done exactly using the same algorithm we use in the population dynamics, with the only difference
that the total number of surveys is constant  in the population dynamics (i.e. it is equal to $L$) and it decreases with 
$I$ on a three
(the number is proportional to $(6\alpha)^{M-I})$). In the limits $L$ and $M$ going to infinity this difference should not be 
relevant.

In order verify if equation (\ref{BOH}) is true is convenient to define a distance $D(t)$ as
\be
D(t)={\sum_{i=1,L}|\vec{s}_{1}(i,t)-\vec{s}_{2}(i,t)|^{2}\over L} \ .
\ee
 \begin{figure} \begin{center}    
      \includegraphics[width=0.6\textwidth]{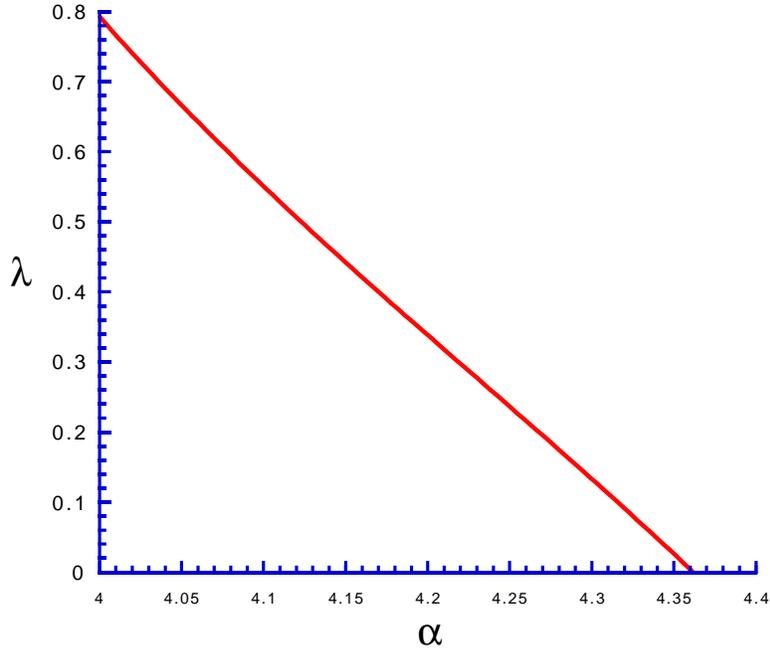}
      \end{center} \caption{ \label{L}
 The numerical results for the exponent $\lambda(\alpha)$, defined in equation (\ref{LA}), as function of $\alpha$.
}\end{figure}

I have numerical studied the properties of  $D(t)$ for large $L$ (up to $L=10^{6}$) finding little $L$ dependence (as 
expected).
I find that for $\alpha<\alpha_{U}\approx 4.36$ for large $t$:
\be
D(t) \propto \exp (-\lambda(\alpha) t) \label{LA}\ ,
\ee
where $\lambda(\alpha)$ is positive for $\alpha<\alpha_{U}$ and vanishes at $\alpha=\alpha_{U}$.  For 
$\alpha>\alpha_{U}$, but not too large both $vec{s}_{1}$ - and $\vec{s}_{2}$ go to the same probability distribution, but 
$D(t) $ does not go to zero.  In other words $-\lambda(\alpha)$ is maximum Liapunov exponent: when it is negative the 
iteration converges to a fixed point, while when it is positive chaos is present.

The estimated value of $\alpha_{U}$, i.e. 4.36, is larger that the value where the complexity vanishes (i.e. 4.27), so that the 
previous conjectures should be correct in the interesting region of positive complexity, where there should be  solutions 
of the satisfiability conditions that correspond to the solutions of the survey propagation equations. 

The reader should 
notice that the survey propagation equations do have a solution also for $\alpha>\alpha_{c}$ and the fact that the 
complexity turns out be negative is a warning that original satisfiability problem does not have any solution. In this 
case the heuristic construction of the surveys is empty.

\section{Conclusions.}
The main propose of this note is the construction (following Aldous \cite{ALDOUS}) of the infinite rooted tree
associated to given satisfiability problem.  This infinite rooted tree plays the role of a bridge among a finite
instance of the problems and the infinite random tree where  analytic computations \cite{MPZ,MZ,MP1,MP2} are done. 
It is argued that existence of an unique stable solution on the infinite tree (that apparently holds for
$\alpha<\alpha_{U}\approx 4.36$) implies the existence of an unique stable solution of the survey equations on a large
system in the same range of $\alpha$.

This result implies that the survey equations that have been used \cite{MPZ,MZ} in an algorithm to find an actual 
solution of an instance of the K-sat problem do have a solution.  However independently from the interest of this 
application of the survey equations, I believe that the conjectures that have been put forward have a mathematical 
interest in their own because they clarify the fundamental hypothesis behind the approach of \cite{MPZ,MZ,MP1,MP2}; 
eventually they can be used to prove similar results in other problems (like the $p$-spin model).

\section*{Acknowledgements}
I thank Marc M\'ezard and Riccardo Zecchina for useful discussions and encouragement.

\end{document}